\documentclass[review]{elsarticle}

\usepackage{lineno,hyperref,latexsym}
\modulolinenumbers[5]

\journal{Journal of \LaTeX\ Templates}

\begin{document}

\begin{frontmatter}

\title{High rate, fast timing Glass RPC for the high $\eta$ CMS muon detectors}
\author[ipnl]{M.\,Gouzevitch\fnref{myfootnote}}
\author[ipnl]{F.\,Lagarde}
\author[ipnl]{I. Laktineh}
\author[ipnl]{V.\,Buridon} 
\author[ipnl]{X.\,Chen}
\author[ipnl]{C.\,Combaret}
\author[ipnl]{A.\,Eynard}
\author[ipnl]{L.\,Germani}
\author[ipnl]{G.\,Grenier}
\author[ipnl]{H.\,Mathez}
\author[ipnl]{L.\,Mirabito}
\author[ipnl]{A.\,Petrukhin}
\author[ipnl]{A.\,Steen}
\author[ipnl]{W.\,Tromeur}
\author[Tsinghua]{\\Y.\,Wang}
\author[Tsinghua]{A.\,Gong}
\author[Omega]{\\N.\,Moreau} 
\author[Omega]{C.\,de la Taille} 
\author[Omega]{F.\,Dulucq}
\author{\\ On behalf of CMS-RPC collaboration} 

\address[ipnl]{Institut de Physique Nucleaire de Lyon, Universite de Lyon, Universite Claude Bernard Lyon 1, CNRS-IN2P3, Villeurbanne, France\\}
\address[Tsinghua]{Tsinghua University, Beijing, China}
\address[Omega]{Omega-\'Ecole Polytechnique, Paris, France}

\fntext[myfootnote]{mgouzevi@cern.ch}

\begin{abstract}
The HL-LHC phase is designed to increase by an order of magnitude the amount of data to be collected by the LHC experiments. To achieve this goal in a reasonable time scale the instantaneous luminosity would also increase by an order of magnitude up to $6 \cdot 10^{34}$\,cm$^{-2}$s$^{-1}$. The region of the forward muon spectrometer
($|\eta| > 1.6$) is not equipped with RPC stations. The increase of the expected particles rate up to 2\,kHz/cm$^2$ ( including a safety factor 3 ) motivates the installation of RPC chambers to guarantee redundancy with the CSC chambers already present. The actual RPC technology of CMS cannot sustain the expected background level. A new generation Glass-RPC (GRPC) using low resistivity glass (LR) is proposed to equip at least the two most far away of the four high eta muon stations of CMS~\cite{upgrade}.
The design of small size prototypes and the studies of their performances under high rate particles flux is presented.
\end{abstract}

\begin{keyword}
\texttt{elsarticle.cls}\sep GRPC \sep RPC \sep timing \sep gaseous detector \sep CMS \sep Upgrade \sep HL-LHC
\end{keyword}

\end{frontmatter}

\begin{figure}[!htb]
\begin{center}
\includegraphics[width=0.9\textwidth]{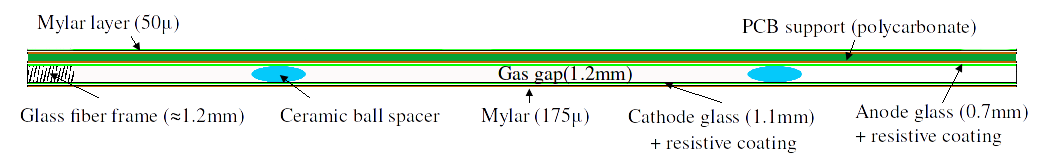}
\includegraphics[width=0.7\textwidth]{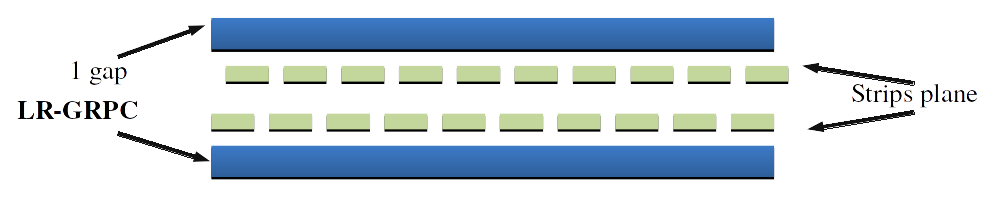}
\caption{Scheme of one small single gap (top) and a schematic of the two single gap GRPC.\label{fig.scheme}}

\end{center}
\end{figure}

The dimensions of the first prototypes are constrained by the largest size of the LR glass plates that could be produced currently: $30\,{\rm cm}\times 32\,{\rm cm}$. Few plates were used to build small prototypes sketched in Fig.~\ref{fig.scheme}.top: a gas gap of $1.2$\,mm separates two $1$\,mm thick LR glass plates (resistivity $10^{10}\,\Omega \cdot$cm) covered with a colloidal graphite coating (surface resistivity of about few M$\Omega/\Box$). Spacers made of glass fiber and ceramic are used to maintain uniform the distance between the two layers. To operate the detectors a gas mixture made of  TFE(93\%), CO2(5\%) and  SF6(2\%) is used~\cite{small}.

The signal was collected using two kinds of PCB electronic plates: the first equipped  with 1\,cm$^2$ pad and readout by 24 64-channel HARDROC ASICs~\cite{hardroc}; the second is equipped with two layers of 128 strips of the same direction and readout with 4 ASICs. The pitch of the strips of each layer is $d=2.5$\,mm with $0.5$\,mm separating two adjacent strips. The strips of one layer are shifted by 1\,mm with respect to those of the other layer in the direction perpendicular to the strips direction (see  Fig.\ref{fig.scheme}.bottom). This configuration, referred to as double-gap, is designed to increase the spatial resolution by looking at the coincidence of fired strips in the two layers. 

The performance of a small GRPC detector was validated for the first time in an electron beam at DESY~\cite{HRate}. More recently the chambers were submitted to a more intensive, wide and energetic muon/pion beam in the CERN-PS line in 2014 and then in CERN-SPS in 2015. 
A telescope built from several small chambers 4 of which are made using LR glass. Additional small chambers made with float glass were also added to compare the behavior of the two kinds of GRPC in high particle rate conditions. The efficiency of the different GRPC were measured using reconstructed straight tracks. For the events where at least 3 layers (strips or pads) were fired a $\chi^2$ fit of a linear track was performed.
The position of the expected hit on each layer was then estimated by extrapolating the track trajectory. The counting of the observed hits within 3.0\,cm around the estimated impact point of the track gave a handle to the efficiency estimation, cluster size and to the strips space-resolution.

\begin{figure}[!htp]
\begin{center}
\includegraphics[width=0.62\textwidth]{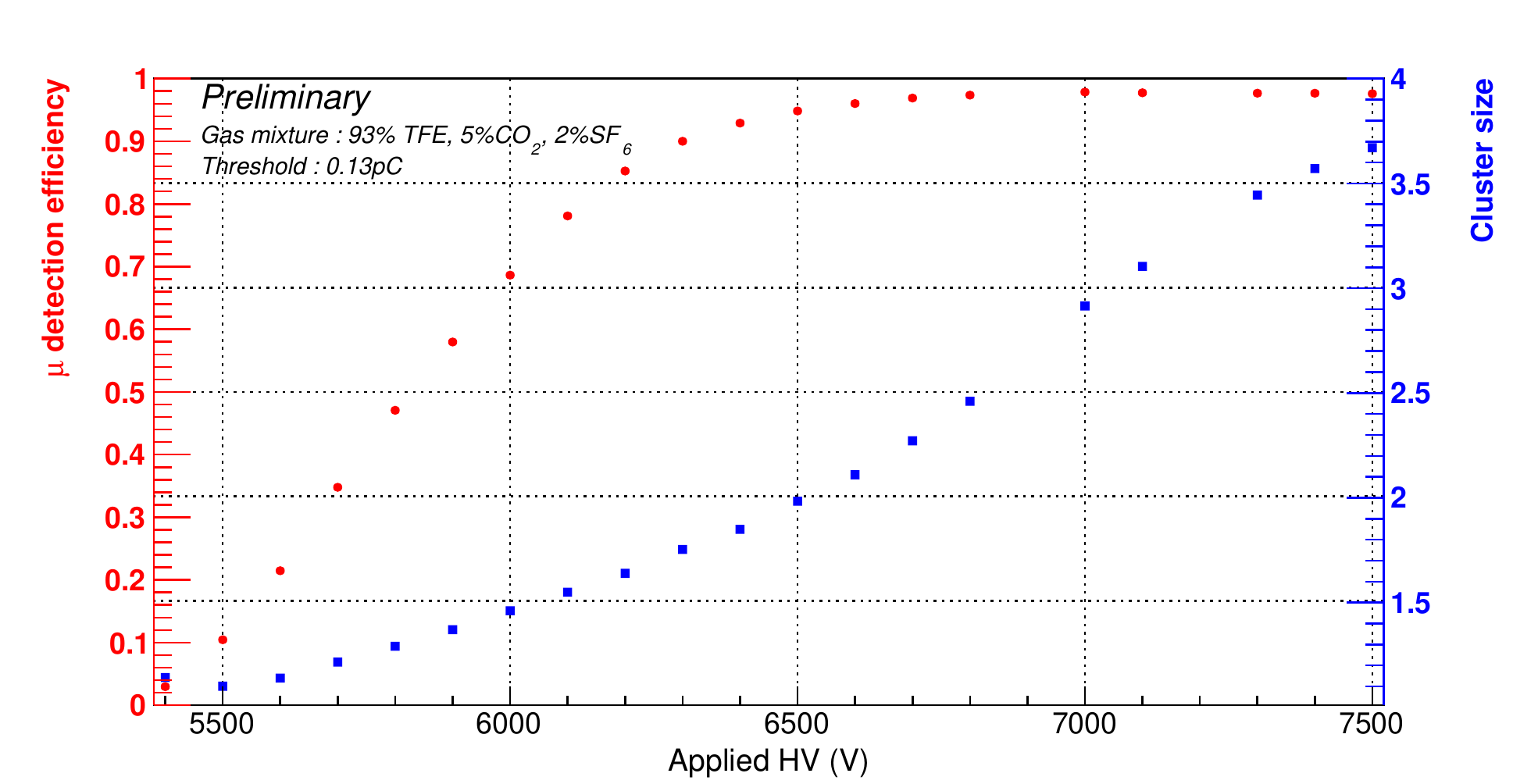}
\includegraphics[width=0.37\textwidth]{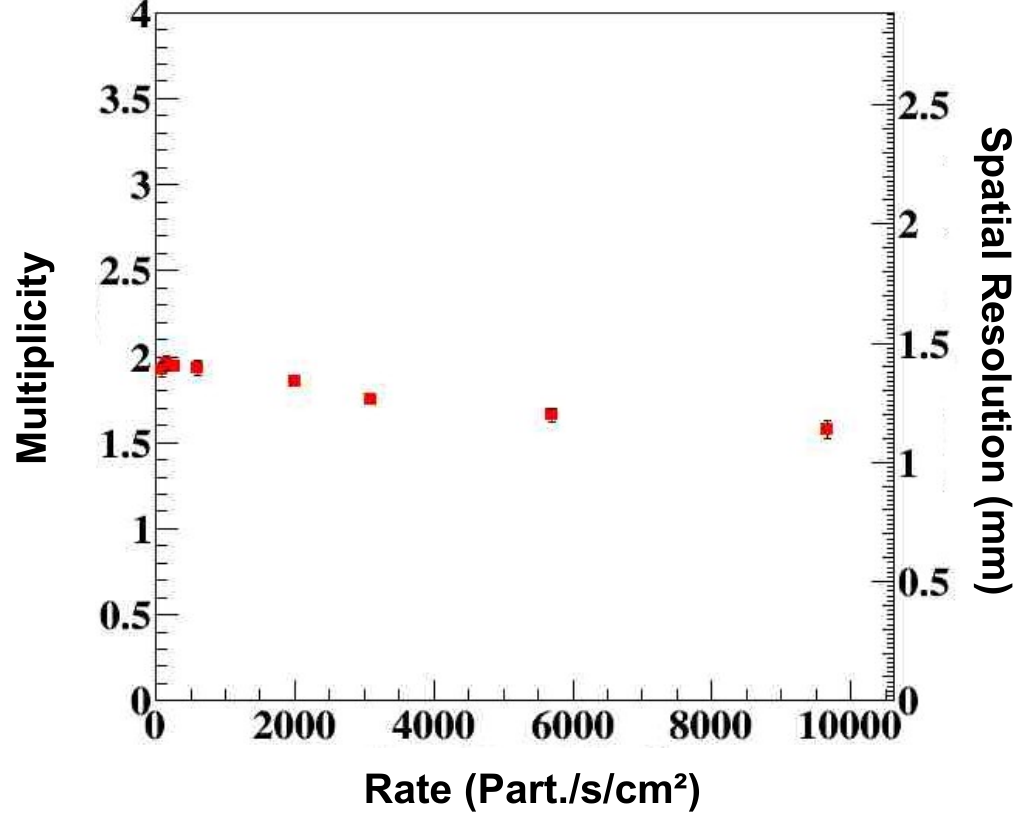}
\caption{Left: Efficiency and cluster size of one single gap GRPC as a function of the applied HV. Right: spatial resolution of double-gap GRPC as function of the particles rate in PS.}
\label{efficiency}
\end{center}
\end{figure}

In Fig.~\ref{efficiency}.left  the evolution of the average efficiency and cluster size of one of the single gap detector with pixel readout are shown as a function of the applied high voltage during the SPS beam test. An efficiency plateau is reached around 6.7\,kV with a relatively small cluster size of 2.2 hits. A similar result was found earlier during the PS test beams for the double-gap configuration. The spatial resolution as function of the rate was estimated by looking at the coincidences of strips in both layers (Fig.~\ref{efficiency}.right). The resolution improves from $1.5$ down to 1\,mm at $10\,\mathrm{kHz/cm^2}$. This trend can be explained by a significant surface charge accumulation at high rate that effectively reduces the cluster size.

\begin{figure}[htp]
\begin{center}
\includegraphics[width=0.95\textwidth]{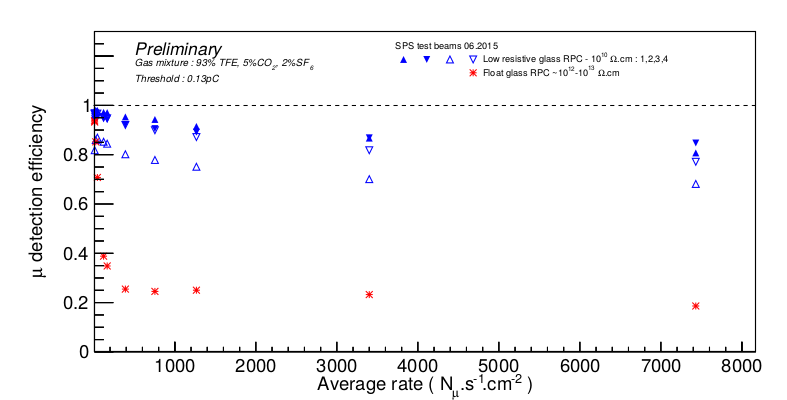}
\caption{The efficiency of the the five detectors as a function of the particle rate in SPS.}
\label{5Chambers}
\end{center}
\end{figure}

The efficiency of the the five detectors as a function of the particle rate are shown in Fig.~\ref{5Chambers}. We observe that at low particle rate all the five chambers are very efficient ( $\epsilon > 90\%$), but the efficiency of the float glass RPC drops dramatically down to $10\%$ when the rate exceeds $0.1\,\mathrm{kHz/cm^2}$ while the four LR chambers keep being efficient at high rate albeit a small efficiency drop.
One LR GRPC exhibits lower efficiency than the three others. We track this inefficiency to the presence of dead channels in the electronic readout that was not corrected when estimating the efficiency. Nevertheless this chamber shows an identical efficiency trend as function of the particles rate up to a normalization factor. 

To conclude we have shown that the LR GRPC is a promising technology that can provide a detection efficiency above 90\% and a spatial resolution of 1\,mm in the radiation conditions expected during HL-LHC program. New tests are ongoing to study the ageing properties and the timing resolution.

\end{document}